\documentclass[preprint2]{aastex}
\usepackage{graphicx}
\usepackage{color}

\begin{document}

\title{Are the Dyson rings around pulsars detectable?}

\author{Z. Osmanov}
\affil{School of Physics, Free University of Tbilisi, 0183, Tbilisi,
Georgia}

\begin{abstract}
In the previous paper \citep{ring} (henceforth Paper-I) we have
extended the idea of Freeman Dyson and have shown that a
supercivilization has to use ring-like megastructures around pulsars
instead of a spherical shell. In this work we reexamine the same
problem in the observational context and we show that facilities of
modern IR telescopes (VLTI and WISE) might efficiently monitor the
nearby zone of the solar system and search for the IR Dyson-rings up
to distances of the order of $0.2$kpc, corresponding to the current
highest achievable angular resolution, $0.001$mas. In this case the
total number of pulsars in the observationally reachable area is
about $64\pm 21$. We show that pulsars from the distance of the
order of $\sim 1$kpc are still visible for WISE as point-like
sources but in order to confirm that the object is the neutron star,
one has to use the UV telescopes, which at this moment
cannot provide enough sensitivity.

\end{abstract}


\section{Introduction}

Kardashev in his well known work \citep{kardashev}, has classified
the extraterrestrial civilization by a technological level they have
achieved. In particular, he introduced three different types of
civilization distinguishing them by means of the ability to consume
energy. According to this classification the level-I is a
civilization similar to ours, thus the one consuming the energy of
the order of $4\times 10^{19}$ergs s$^{-1}$. Level-II corresponds to
a civilization consuming almost the total energy of their host star
- $4\times 10^{33}$ergs s$^{-1}$ and Level-III is a civilization
harnessing almost the total energy of its own galaxy: $4\times
10^{44}$ergs s$^{-1}$.

A couple of years earlier before publishing the paper of Kardashev,
the prominent physicist Freeman Dyson has suggested that if such
superadvanced (in the terminology of Kardashev, Level-II)
extraterrestrials exist, for increasing efficiency of energy
consumption they can construct a thin spherical shell with radius
$\sim 1$AU surrounding a host star \citep{dyson}. It has been argued
that for such distances the sphere will be in the so-called
habitable zone (HZ) and therefore the sphere will have the
temperature of the order of $(200-300)$K, making this object visible
in the infrared spectrum.

An interest to the search for the infrared Dyson-like sources has
been increased after the detection of an enigmatic object,
KIC8462852, discovered by the Keppler mission \citep{kic846}. It has
been shown that the flux of the mentioned object has been
characterized by aperiodic dips of the order of $20\%$. On the other
hand the authors have confirmed that the irregularities might have
not been caused by any instrumental or data processing factors. To
investigate the hypothesis that strange behavior of KIC 8462852 (and
some other objects: KIC 12557548, CoRoT-29) is caused by artificial Dyson-like cosmic
structures a series of works has been performed
\citep{wright,opt,radio}, but the main question concerning the
origin of the unrealistically high level of flux dip still remains
open. Whatever this origin is the discovery of KIC 8462852 has
revived a search for artificial cosmic megastructures which has been
started in the first decade of this century
\citep{carrigan,jugaku,timofeev,slish}.

\begin{figure}
  \centering {\includegraphics[width=7cm]{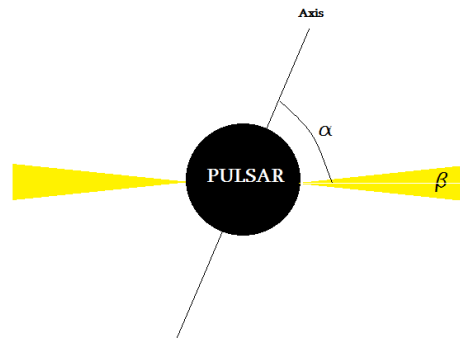}}
  \caption{Here we schematically show the pulsar, its axis of rotation,
  and two emission channels with an opening angle $\beta$.}\label{fig1}
\end{figure}

On the other hand, it is clear that such cosmic megastructures
require enormous material to construct them. For avoiding this
difficulty, recently we have proposed a certain extension of Dyson's
idea. In particular, in Paper-I  we have examined the Level-II 
civilisation, which colonised a nearby area of a pulsar emitting electromagnetic radiation in narrow
channels (see Figure \ref{fig1}). Therefore, it means that 
the considered scenario should be examined as a probable one. Then, unlike the Dyson spheres one
has to construct ring-like structures. In particular, if the
inclination angle, $\alpha$, between the axis of rotation and the
emission channels is close to $90^o$, the ring should be located in
the equatorial plane and as it has been shown  \citep{ring} for
relatively slowly rotating neutron stars the sizes of the rings
could vary from $10^{-4}$AU to $10^{-1}$AU respectively. As a
result, such a megastructure would require less material than in
case of spheres. It is clear that such rings might be significantly
perturbed by tidal stresses in terms of strong centrifugal outflows and radiation \citep{guda}.
Therefore, they cannot be stable and sooner or later gravitation
will inevitably destabilise the construction. We have analysed
dynamics of destabilisation and found that power required to restore
stability is much less than power received from the pulsar.
Therefore, the search for the ring-like megastructures around slowly
spinning pulsars seen in the infrared spectrum might be quite
promising. As to the rapidly rotating pulsars, they are very
powerful and harvesting their energy would be quite profitable, but
a habitable zone would be much farther and mass of a material
required for construction the mega-ring would exceed the total mass
of all planets, asteroids, comets, centaurs and interplanetary dust
in a typical planetary system by several orders of magnitude.
Therefore in this paper we focus on the megastructures around normal
pulsars and study the possibility of detecting such cosmic
constructions in a relatively nearby zone of the solar system. For
this purpose we discuss the possible requirements telescopes have to
satisfy.

The organization of the paper is the following: after outlining the
major results of Paper-I, we consider the observational features of
the Dyson rings and make necessary calculations in Sec. 2 and in
Sec. 3 we summarise our results.

\section[]{Previous results and discussion}

In this section we outline the results obtained in Paper-I, estimate
characteristic parameters of the cosmic megastructures and study the
possibility of their detection by means of a set of observations.

\subsection[]{Outline of the results of Paper-I}

It is well known that pulsars, having enormous energy accumulated in
rotation gradually slow down their rotation rate and as a result
loose energy, with a certain fraction, $\kappa$, going to emission
with the following luminosity \citep{ring}
$$L\approx 3.8\times 10^{31}\times
\left(\frac{\kappa}{0.1}\right)\times\left(\frac{0.5s}{P}\right)^3\times\left(\frac{\dot{P}}{10^{-15}ss^{-1}}
\right)\times$$
\begin{equation}
\label{lumin} \times\left(\frac{M}{1.5M_{\odot}}\right)ergs\ s^{-1},
\end{equation}
where $P$ is the rotation period of the pulsar, $\dot{P}\equiv
dP/dt>0$, $M\approx 1.5\times M_{\odot}$ and $M_{\odot}\approx
2\times 10^{33}$g are the pulsar's mass and the solar mass
respectively and we normalize the period on an average value of the
most probable one \citep{catalogue}.

In Paper-I we have implied the definition of the HZ as a region
irradiated by the same energy flux as the Earth \citep{hanslmeier}.
Then, it is straightforward to show that radius of the HZ is given
by
$$R_{_{HZ}}\approx 0.1\times\left(\frac{\kappa}{0.1}\right)^{1/2}\times
\left(\frac{0.5s}{P}\right)^{3/2}\times$$
\begin{equation}
\label{rHZ} \times\left(\frac{\dot{P}}{10^{-15}ss^{-1}}
\right)^{1/2}\times\left(\frac{M}{1.5M_{\odot}}\right)^{1/2}AU.
\end{equation}

For estimating the effective area of the construction one has to
take into account the opening angle of the emission channel
\citep{rudsuth}
\begin{equation}
\label{angle} \beta\approx \frac{\pi}{3}\times
\left(\frac{0.5}{P}\right)^{13/21}\times\left(\frac{\dot{P}}{10^{-15}ss^{-1}}
\right)^{1/14},
\end{equation}
automatically defining the effective area of the ring having the
chape of the spherical segment \citep{ring}
\begin{equation}
\label{area} A_{ef}\approx 8\pi R_{_{HZ}}^2\sin\left(\beta/2\right),
\end{equation}
where for simplicity we have assumed that $\alpha\approx\pi/2$. In
the Paper-I we have obtained an expression of the ring's temperature
\begin{equation}
\label{balance} T =
\left(\frac{L}{A_{_{ef}}\sigma}\right)^{1/4}\approx 390K,
\end{equation}
where $\sigma\approx 5.67\times 10^{-5}$erg/(cm$^2$K$^4$) is the
Stefan-Boltzmann constant.

\subsection[]{Observational features of Dyson-rings}

By combining Eq. (\ref{balance}) with the Wien's law
\begin{equation}
\label{wien} \lambda_m = \frac{b}{T}\approx 7.4\times 10^3nm,
\end{equation}
it is evident that the cosmic megastructure around a pulsar will be
visible in the infrared spectrum. The Very Large Telescope
Interferometer (VLTI) could be a quite promising instrument for
observing the rings. In particular, the VLTI's angular sensitivity
is $0.001$mas (milliarcsecond) (see the technical characteristics of
the VLTI
\footnote{https://www.eso.org/sci/facilities/paranal/telescopes/vlti}).
By implying the latter, one can straightforwardly show that the
maximum distance, where the ring with the diameter $2R_{_{HZ}}$
still will be observable by the VLTI is expressed as
\begin{equation}
\label{dmax} r_{max}\approx\frac{2R_{_{HZ}}}{\theta}\approx 0.2kpc.
\end{equation}
On this distance the infrared spectral flux density (power per unit
of area for the unit interval of frequency) is approximately given
by
$$F^{_{IR}}_{\nu}\equiv\frac{dE}{dtdAd\nu}\approx
\frac{L}{4\pi r^2\nu_{_{IR}}}\approx$$
$$\approx 7.4\times
10^{-26}\left(\frac{0.2kpc}{r}\right)^2
ergs\;s^{-1}\;cm^{-2}\;Hz^{-1}=$$
\begin{equation}
\label{fluxIR}  = 7.4\left(\frac{0.2kpc}{r}\right)^2 mJy,
\end{equation}
where $\nu_{_{IR}} = c/\lambda$ and we have assumed that almost the total energy of 
a pulsar is radiated by the Dyson ring in the infrared spectrum. The derived
spectral flux density can be detected by the VLTI. It is worth
noting that the pulsars are distributed mostly in the galactic plane
and their surface density, $\rho$, in the local area of galaxy is of
the order of $[520\pm 170]$kpc$^{-2}$ \citep{manch}. Therefore, it
is expected that when monitoring cylindrical volume in the galactic
plane corresponding to the circular area with radius $r_{max}$ one
can expect in total the following number of pulsars
\begin{equation}
\label{numb1} N_{0.2}\approx \rho\pi r_{max}^2\approx 64\mp 21,
\end{equation}
where the subscript $0.2$ means that the number is calculated for
the distance $0.2$kpc. At this moment this is the maximum number of
pulsars, among which one can search for the potential infrared
sources in the nearby zone of the Solar system. It is clear that the
most distant objects observed by the VLTI will not be seen in
detail.

The Wide-field Infrared Survey Explorer (WISE) might be very useful,
although, since the angular resolution is not very high, one can use
this instrument for searching the point like sources, because the
sensitivity might be better than $0.1$mJy
\footnote{http://wise2.ipac.caltech.edu/docs/release/allsky}. The
infrared Dyson-rings observed from the distance $r=1$kpc will have
the flux density of the order of $0.3$mJy (see Eq. \ref{fluxIR}).
This in turn means that the number of expected pulsars is much
higher than for $0.2$kpc
\begin{equation}
\label{numb} N\approx \rho\pi r^2\approx 1600\pm 530.
\end{equation}
The infrared Dyson-rings probably might have another interesting
observational feature. In particular, it has been shown by
\cite{zhang} that for the normal pulsars with $P\sim 0.5$s and
$\dot{P}\sim 10^{-15}$ss$^{-1}$, having the age of the order of
$\tau\equiv P/2\dot{P}\sim 1.6\times 10^7$yr, the surface
temperature is given by
\begin{equation}
\label{T} T(\tau)\approx 2.8\times
10^5K\left(\frac{10^6yr}{\tau}\right)^{0.5}\approx 7\times 10^4K,
\end{equation}
which corresponds to the UV spectral band, $6$eV. This means, that
the pulsar encircled by the ring seen in the infrared spectrum
potentially might be detected in the UV as well. This additional
observation might be useful for distances less than $0.2$kpc, but
for higher distances it will be even more significant, because since
the infrared source is point like one needs another confirmation
(may be not necessarily enough) that in the same location there is a
neutron star.

The corresponding UV total luminosity writes as
\begin{equation}
\label{LUV} L_{_{UV}}\approx 4\pi\sigma
R_{_{NS}}^2T_{_{UV}}^4\approx 1.7\times 10^{28} ergs\;s^{-1},
\end{equation}
where $T_{UV}\approx 7\times 10^4$K is the neutron star's surface
temperature and $R_{_{NS}}\approx 10$km is the radius of the star.
The spectral flux density, for the distance $0.2$kpc then can be
estimated as
$$F^{_{_{UV}}}_{\nu}\equiv\frac{dE}{dtdAd\nu}\approx
\frac{L_{_{UV}}}{4\pi r^2\nu_{_{UV}}}\approx $$
\begin{equation}
\label{fluxUV} \approx 2.5\times
10^{-30}\left(\frac{0.2kpc}{r}\right)^2
ergs\;s^{-1}\;cm^{-2}\;Hz^{-1},
\end{equation}
where $\nu_{_{UV}}\approx 1.5\times 10^{15}$Hz is the leading
frequency of emission. As it is clear from this estimate, the flux
is quite low for distances of the order of $0.2$kpc and will be even
less for larger values of $r$.

\section{Conclusion}

We outlined the main results of Paper-I where the extension of the
idea of Freeman Dyson has been considered and it has been shown that
for relatively slowly rotating pulsars with $P=0.5$s and $\dot{P} =
10^{-15}$ss$^{-1}$ the habitable zone will be on distances of the
order of $0.1$AU. By examining the ring in the habitable zone, we
have found that the temperature of this megastructure will be of the
order $390$K, indicating that the Dyson megastructure will be visible in the
IR band.

We have considered the  sensitivity of VLTI and by taking into account its higher
possible angular resolution, $0.001$mas, it has been shown that the
maximum distance $\sim 0.2$kpc leads to the IR spectral density of
the order of $7.4$mJy, which in turn, can be detected by the VLTI.
We have argued that by monitoring the nearby zone of the solar
system approximately $64$ pulsars are expected to be located inside
it.

It has been shown that in order to observe the infrared sources up to
the distances of the order of $\sim$kpc, the IR flux density is of
the order of $3\times 10^{-27}$ ergs s$^{-1}$ cm$^{-2}$ Hz$^{-1}$,
which can be detected by the IR telescope WISE. On the other hand,
the angular resolution is not enough to see the structure of the
ring. Therefore, it is necessary to find another feature of the
distant IR ring.

By implying the model developed in \citep{zhang} we have found that
the temperature of the neutron star's surface is $7000$K, which in
the electromagnetic spectrum corresponds to the UV band. The
corresponding spectral energy flux density for the distance $\sim
1$kpc, $10^{-31}$ ergs s$^{-1}$ cm$^{-2}$ Hz$^{-1}$ is so small that
at this moment no UV instrument has such a high sensitivity.

As we see, the search of infrared rings is quite promising for
distances up to $\sim 0.2$kpc, where one will be able to monitor
potentially $64\pm 21$ pulsars by using the IR instruments.
Observation of distant pulsars (up to $\sim 1$kpc), although will
significantly increase the total number of potential objects - to
$1600\pm 530$, but at this moment the UV instruments cannot provide
such a level of sensitivity.

\section*{Acknowledgments}
The author would like to thank a referee for valuable comments. The research was supported by the Shota Rustaveli National Science Foundation grant (DI-2016-14). 

\begin{thebibliography}{99}
\bibitem[Boyajian et al.(2016)]{kic846} Boyajian, T.S. et al., 2016,
MNRAS, 457, 3988
\bibitem[Carrigan(2009)]{carrigan} Carrigan, R. A., 2009, ApJ, 698, 2075
\bibitem[Dyson(1960)]{dyson} Dyson, F., 1960,
Science, 131, 1667
\bibitem[Gudavadze et al.(2015)]{guda} Gudavadze, I., Osmanov, Z. \&
Rogava, A., 2015, Int. J. Mod. Phys. D., 24, 1550042
\bibitem[Hanslmeier(2009)]{hanslmeier} Hanslmeier, A., 2009, Habitability
and Cosmic Catastrophes, Springer-Verlag (Berlin Heidelberg), 2009

\bibitem[Harp et al.(2016)]{radio} Harp, G. R., Richards, J., Shostak, S.,
Tarter, J. C., Vakoch, D. A. \& Munson, C., 2016, ApJ, 825, 1
\bibitem[Jugaku \& Nishimura(2002)]{jugaku} Jugaku, J. \&
Nishimura, S., 2002, in Proc. IAU Symp. 213, Bioastronomy 2002: Life
Among the Stars, ed. R. Norris \& F. Stootman (San Francisco, CA:
ASP), 437
\bibitem[Kardashev(1964)]{kardashev} Kardashev, N. S., 1964,
AJ, 8, 217
\bibitem[Manchester(1979)]{manch} Manchester, N. S., 1979, Aust. J. Phys., 32, 1
\bibitem[Manchester et al.(2005)]{catalogue} Manchester, N. S., Hobbs, G. B.,
Teoh, A. \& Hobbs, M., 2005, AJ, 129, 1993
\bibitem[Osmanov(2016)]{ring} Osmanov, Z., 2016, IJAsB, 15, 127
\bibitem[Ruderman \& Sutherland(1975)]{rudsuth} Ruderman,
M. A. \& Sutherland, P. G., 1975, ApJ, 196, 51
\bibitem[Schuetz et al.(2016)]{opt} Schuetz, M., Vakoch, D. A., Shostak, S.
\& Richards, J., 2016, ApJL, 825, 1
\bibitem[Slish(1985)]{slish} Slish, V. I., 1985, in The Search for
Extraterrestrial Life: Recent Developments, ed. M. D. Papagiannis
(Boston, MA: Reidel Pub. Co.), 315
\bibitem[Timofeev et al.(2000)]{timofeev} Timofeev, M. Y., Kardashev, N. S. \& Promyslov, V. G.,
2000, Acta Astronautica J., 46, 655
\bibitem[Wright et al.(2016)]{wright} Wright, J. T., Cartie, K. M., Zhao, M.,
Jontof-Hunter, D. \& Ford, E. B., 2016, ApJ, 816, 22
\bibitem[Zhang \& Harding(2000)]{zhang} Zhang, B. \& Harding, A.K.,
2000, ApJ, 532, 1150









\end{thebibliography}
\end{document}